\documentclass[twocolumn,secnumarabic,amssymb, nobibnotes, aps, prl, superscriptaddress]{revtex4-2}

\usepackage{graphicx}
\usepackage{natbib}
\usepackage{xcolor}
\usepackage[colorlinks,allcolors=blue]{hyperref}
\usepackage{amsmath}
\usepackage{ulem}
\usepackage{lipsum}

\begin{document}

\title{Chiral emission induced by optical Zeeman effect in polariton micropillars}

\author{B.~Real}
\email{bastian.realelgueda@univ-lille.fr}
\affiliation{Univ. Lille, CNRS, UMR 8523 -- PhLAM -- Physique des Lasers Atomes et Mol\'ecules, F-59000 Lille, France}

\author{N.~Carlon Zambon}
\affiliation{Universit\'e Paris-Saclay, CNRS, Centre de Nanosciences et de Nanotechnologies, 91120, Palaiseau, France}

\author{P.~St-Jean}
\affiliation{Universit\'e Paris-Saclay, CNRS, Centre de Nanosciences et de Nanotechnologies, 91120, Palaiseau, France}

\author{I.~Sagnes}
\affiliation{Universit\'e Paris-Saclay, CNRS, Centre de Nanosciences et de Nanotechnologies, 91120, Palaiseau, France}

\author{A.~Lema\^itre}
\affiliation{Universit\'e Paris-Saclay, CNRS, Centre de Nanosciences et de Nanotechnologies, 91120, Palaiseau, France}

\author{L.~Le Gratiet}
\affiliation{Universit\'e Paris-Saclay, CNRS, Centre de Nanosciences et de Nanotechnologies, 91120, Palaiseau, France}

\author{A.~Harouri}
\affiliation{Universit\'e Paris-Saclay, CNRS, Centre de Nanosciences et de Nanotechnologies, 91120, Palaiseau, France}

\author{S.~Ravets}
\affiliation{Universit\'e Paris-Saclay, CNRS, Centre de Nanosciences et de Nanotechnologies, 91120, Palaiseau, France}

\author{J.~Bloch}
\affiliation{Universit\'e Paris-Saclay, CNRS, Centre de Nanosciences et de Nanotechnologies, 91120, Palaiseau, France}

\author{A.~Amo}
\affiliation{Univ. Lille, CNRS, UMR 8523 -- PhLAM -- Physique des Lasers Atomes et Mol\'ecules, F-59000 Lille, France}

\begin{abstract} 

The low sensitivity of photons to external magnetic fields is one of the major challenges for the engineering of photonic lattices with broken time-reversal symmetry. Here we show that time-reversal symmetry can be broken for microcavity polaritons in the absence of any external magnetic field thanks to polarization dependent polariton interactions. Circularly polarized excitation of carriers in a micropillar induces a Zeeman-like energy splitting between polaritons of opposite polarizations. In combination with optical spin-orbit coupling inherent to semiconductor microstructures, the interaction induced Zeeman splitting results in emission of vortical beams with a well-defined chirality. Our experimental findings can be extended to lattices of coupled micropillars opening the possibility of controling optically the topological properties of polariton Chern insulators.

\end{abstract}
\date{\today}%
\maketitle

Lattices of coupled waveguides and photonic resonators are some the most remarkable platforms to explore condensed-matter phenomena in the optics realm~\cite{Christodoulides2003}, including topological phases of matter~\cite{Ozawa2019}. The possibility of engineering the hopping strength, the geometry and the onsite energy in a very precise way are some of the assets of these systems to explore wavepacket dynamics in periodic~\cite{Christodoulides2003} and non-periodic structures~\cite{Zilberberg2021} 
using standard optical techniques.

One of the major challenges when engineering photonic lattices is the implementation of hamiltonians with broken time-reversal symmetry. This ingredient in the photonic toolbox allows the study of photonic Chern insulators and non-reciprocal structures. A natural way of breaking time-reversal symmetry is the application of an external magnetic field. For instance, the first demonstration of a Chern insulator for electromagnetic waves was realised in the microwave domain using a lattice made of a gyrotropic material in subject to a magnetic field~\cite{Wang2009}. Later, emission from chiral edge modes was reported at telecom and near-infrared wavelengths in YIG photonic crystals~\cite{Bahari2017, Bahari2021} and in a lattice of coupled polariton micropillars~\cite{Klembt2018}, respectively. However, the relatively low magnetic susceptibility of these systems resulted in topological gaps of limited size. Moreover, a very interesting perspective would be the control of the magnitude and direction of the magnetic field at the micron scale in such photonic lattices. This would allow the exploration of lattice hamiltonians subject to, for instance, staggered magnetic fields.

Semiconductor microcavities are an excellent platform to overcome these challenges in the near infrared. Their eigenmodes are microcavity polaritons, which arise from the strong coupling between photons confined in a semiconductor microcavity and excitonic excitations of a quantum well embedded in the structure~\cite{Carusotto2013}. The excitonic component provides matter-like properties to this photonic quasiparticles. For instance, under an external magnetic field, polaritons in GaAs-based microcavities show a Zeeman splitting inherited from the excitonic Zeeman splitting of GaAs quantum wells~\cite{Larionov2010, Walker2011, Pietka2015, Sturm2015}. Magnetic fields have  been as well used to engineer gauge potentials~\cite{Lim2017}. The excitonic component results also in significant polariton-polariton interactions. They are dominated by excitonic exchange terms~\cite{Ciuti1998b} giving rise to a strong anisotropy depending on the polariton polarization~\cite{Shelykh2004c}: polaritons of same circular polarization (i.e., spin) interact much stronger than polaritons of opposite polarizations~\cite{Vladimirova2010, Sarkar2010, Adrados2010, Sich2014}. Interestingly, polarization dependent interactions also take place between polariton modes and the reservoir of excitons excited under non-resonant pumping~\cite{Cilibrizzi2015,pickup_optical_2018, Gnusov2020}. Therefore, by optically injecting a spin polarized exciton gas in the quantum wells, it is possible to induce different blueshifts for polaritons of same and opposite polarizations than that of the exciton gas, i.e., an interaction induced Zeeman splitting. Recently, this effect has been reported in transition metal dichalcogenides~\cite{Li2021}. As the exciton reservoir can be localised in a few microns scale~\cite{Ferrier2011}, the interaction induced Zeeman splitting could be engineered with onsite precission in a polariton lattice.

\begin{figure}[t!]
\begin{center}
  \includegraphics[width=0.47\textwidth]{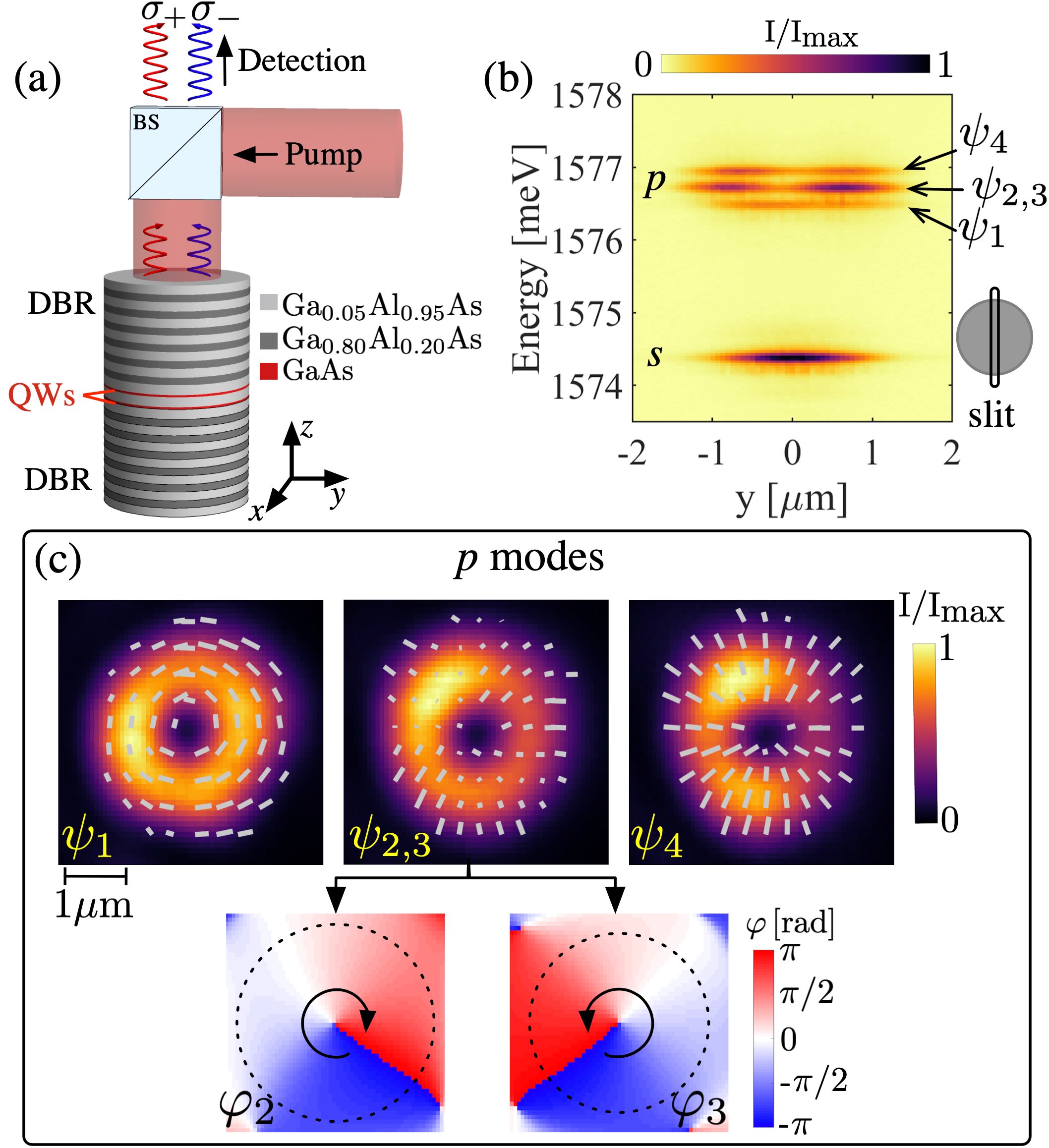}
  	\caption{\label{fig1}
  	(a) Sketch of the micropillar and pumping setup. (b) Photoluminescence spectrum of a $2.8$-$\mu$m-diameter micropillar, showing \textit{s} and \textit{p} orbital modes. (c) Top, real-space emission measured at low pump irradiance ($0.26$~kW~cm$^-2$) at the energy of  $\psi_{1,2,3,4}$ modes indicated in (b). The bars show the measured orientation and degree of linear polarisation at each point in space. 
  	Bottom, retrieved phase at the energy of the $\psi_2$ and  $\psi_3$ modes when selecting $\sigma_+$ (left) and $\sigma_-$ (right) circular polarization of emission.
	}
\end{center}
\end{figure}

In this work we demonstrate the breaking of time-reversal symmetry for polaritons in a GaAs-based  micropillar without the need of any external magnetic field. To do so we create an optical Zeeman splitting for polaritons by injecting a gas of spin polarized excitons. Moreover, in combination with optical spin-orbit coupling inherent to semiconductor microstructures~\cite{Kavokin2004,Sala2015, Dufferwiel2015}, the  optically induced Zeeman splitting results in emission of vortical beams with a well-defined chirality. The investigated micropillar structure is the building block to fabricate polariton lattices~\cite{Jacqmin2014}: our experimental findings establish a first step towards the optical control of topological Chern phases expected to appear in lattices of coupled polariton micropillars combining a Zeeman effect and spin-orbit coupling~\cite{Nalitov2014b,Karzig2015,Bleu2017b}.

For our experiments we use a semiconductor micropillar with a diameter of $2.8\,\mu$m [Fig.~\ref{fig1}(a)].  It is etched from a $\lambda/2$ Ga$_{0.05}$Al$_{0.95}$As microcavity ($\lambda=780$~nm) embedded in two distributed Bragg reflectors of 40 and 28 pairs of $\lambda/4$ layers of Ga$_{0.05}$Al$_{0.95}$As/Ga$_{0.80}$Al$_{0.20}$, where $\lambda$ is the wavelength of the fundamental Fabry-Perot mode of the cavity. Three sets of four 7 nm GaAs quantum wells (QWs) are grown inside the microcavity at the three central maxima of the electromagnetic field. At the cryogenic temperature of the experiments ($5$~K), cavity photons and QW excitons enter the strong-coupling regime giving rise to exciton polaritons with a measured Rabi splitting of $15$~meV. 

To probe the energy spectrum of the micropillar, we carry out photoluminescence experiments using a Ti:Sapph continuous-wave laser tuned out of resonance ($\lambda=744.4$ nm). This pump laser is tightly focused on the micropillar by using an aspheric lens with 8~mm focal length (NA$=0.5$), producing a spot with a full width at half maximum (FWHM) of $2.5\,\mu$m. All experiments are performed in reflection geometry, as sketched in Fig.~\ref{fig1}(a). The photoluminescence is resolved both in energy, space and polarization using an imaging spectrometer coupled to a CCD camera.

Figure~\ref{fig1}(b) displays the low power photoluminescence spectrum measured along the vertical diameter of a micropillar with a photon-exciton detuning $\delta_{cx}\approx-7.8$~meV for the lowest energy mode (excitonic fraction of $\sim 27\%$). The excitation beam is circularly polarized and the emission is detected in the same polarization. The two sets of energy levels displayed in the panel correspond to the fundamental and first-excited modes, called \textit{s} and \textit{p}, respectively, owing to their symmetry. The eigenstates of the micropillar can be accurately described in a basis of Laguerre-Gauss modes in cylindrical coordinates ($r,\theta$) and polarization pseudospins~\cite{Gerard1996,Reithmaier1997,Dufferwiel2015}: $\psi_{n,l}^{\sigma}=C_{n,l}^\sigma(r) e^{il\theta}$, where $C_{n,l}(r)$ is the radial part of the Laguerre-Gauss modes. This basis is characterised by three quantum numbers: the radial quantum number $n$, which selects the $s$ or $p$ modes, the orbital angular momentum (OAM) $l$, and the circular polarization $\sigma_{\pm}$. The two \textit{s} modes present a OAM of $l=0$ and are degenerate in polarization ($\sigma_{\pm}$) at $E_s=1574.12$~meV.

The splitting of the triplet of $p$ modes visible in Figure~\ref{fig1}(b) at around $E_p=1576.74
$~meV is a consequence of the optical spin-orbit coupling present in dielectric microcavities. In planar structures it manifests in the form of a linear polarization splitting between modes polarized along and perpendicular to the propagation direction within the cavity~\cite{Panzarini1999b, Kavokin2005, Leyder2007}. In microstructured microcavities, the spin-orbit coupling gives rise to a fine structure of modes that mix the orbital and polarization degrees of freedom. This was first reported in a hexagonal molecule of coupled micropillars in Refs.~\cite{Sala2015, CarlonZambon2019}, whose $|l|=1$ multiplet shows a triplet similar to the one observed in Fig.~\ref{fig1}(b). The exact same fine structure was found for \textit{p} modes in a single photonic dot open cavity by Dufferwiel and coworkers, who described the spin-orbit coupling effect in these modes with the following Hamiltonian~\cite{Dufferwiel2015}: 

\begin{equation}
H_p=
\begin{pmatrix}
E_p-\frac{\Delta E_z}{2}&0&0&0\\
0&E_p+\frac{\Delta E_z}{2}&t_{\text{SOC}}&0\\
0&t_{\text{SOC}}&E_p-\frac{\Delta E_z}{2}&0\\
0&0&0&E_p+\frac{\Delta E_z}{2}
\end{pmatrix}\,,
\label{Hami2}
\end{equation}

\noindent which is written in the basis $\ |l,\sigma\rangle$: $\{ |+1,\sigma_+\rangle,|+1,\sigma_-\rangle,|-1,\sigma_+\rangle,|-1,\sigma_-\rangle \}$. $E_p$ is the energy of the \textit{p} modes in the absence of spin-orbit coupling and Zeeman effect, $\Delta E_z$ is the optical Zeeman splitting (which we will introduce later) and $t_{\text{SOC}}$ is the strength of the spin-orbit coupling. When $\Delta E_z=0$ the four modes present the following eigenstate fine structure:
\begin{align}
|\psi_1\rangle&=\frac{1}{\sqrt{2}}\left(|-1,\sigma_+\rangle-|+1,\sigma_-\rangle\right)\,,\nonumber \\
|\psi_2\rangle&=|+1,\sigma_+\rangle\nonumber \,,\\   |\psi_3\rangle&=|-1,\sigma_-\rangle\nonumber\,, \\
|\psi_4\rangle&=\frac{1}{\sqrt{2}}\left(|-1,\sigma_+\rangle+|+1,\sigma_-\rangle\right)\,,
\label{pmodes}
\end{align}

\noindent with energy $E_1=E_{\textit{p}}- t_{\text{SOC}}$, $E_2=E_3=E_{\textit{p}}$ and $E_4=E_{\textit{p}}+t_{\text{SOC}}$. This simple model explains the three \textit{p} emission lines visible in Fig.~\ref{fig1}(b) at low excitation power, with $t_{\text{SOC}}=254\pm17\,\mu$eV.

The combination of orbital and polarization modes of the eigenstates results in azimuthal and radial polarization textures for the lowest ($\psi_1$) and highest ($\psi_4$) energy states, respectively. Figure~\ref{fig1}(c) shows the measured direction and degree of the linear polarization, represented by the orientation and length of the bars, extracted from polarization-resolved measurements for the three emission energies of the \textit{p} multiplet~\cite{Supplementary}. The lowest and highest energy levels present, indeed, azimuthal and radial linear-polarization textures, as reported in Refs.~\cite{Sala2015, Dufferwiel2015}. 

The two middle states ($\psi_2$ and $\psi_3$) are degenerated at energy $E_p$. Each one presents a phase vorticity of topological charge $\pm 1$ entangled to a circular polarization $\sigma_{\pm}$, respectively. The observed polarization pattern at the energy of these modes [Fig.~\ref{fig1}(c) top panel] is determined by a phase relation between the two modes established by local disorder~\cite{Dufferwiel2015}. The presence of $l=\pm1$ OAM coupled to $\sigma_{\pm}$ circular polarizations in states $\psi_2$ and $\psi_3$ can be evidenced by doing a polarization-resolved interferometric measurement. As a phase reference we use the emission from a region of the micropillar of less than 1~$\mu$m$^2$, which is enlarged and combined with the image of the micropillar at the entrance slit of the spectrometer. We follow the procedure described in Ref.~\cite{Supplementary}. The retrieved phase pattern at the energy of $\psi_2$ and $\psi_3$ for $\sigma_+$ and $\sigma_-$ polarizations of emission, respectively, is shown in the lower panels of Fig.~\ref{fig1}(c). They display clear $2\pi$  and $-2\pi$ windings of the phase in the micropillar. 

One of the most interesting features of this level structure is that it can give rise to chiral modes with non-zero orbital winding if a polarization splitting is induced between modes $\psi_2$ and $\psi_3$. Without inducing any splitting, Carlon Zambon and co-workers showed lasing in one of these degenerated modes in the weak coupling regime using a hexagonal molecule of coupled micropillars~\cite{CarlonZambon2019}. Here we will show that a splitting between these levels can indeed be induced taking advantage of polarization dependent interactions. In this way time reversal symmetry is broken without the need of any external magnetic field.

\begin{figure}[t!]
\begin{center}
  \includegraphics[width=0.5\textwidth]{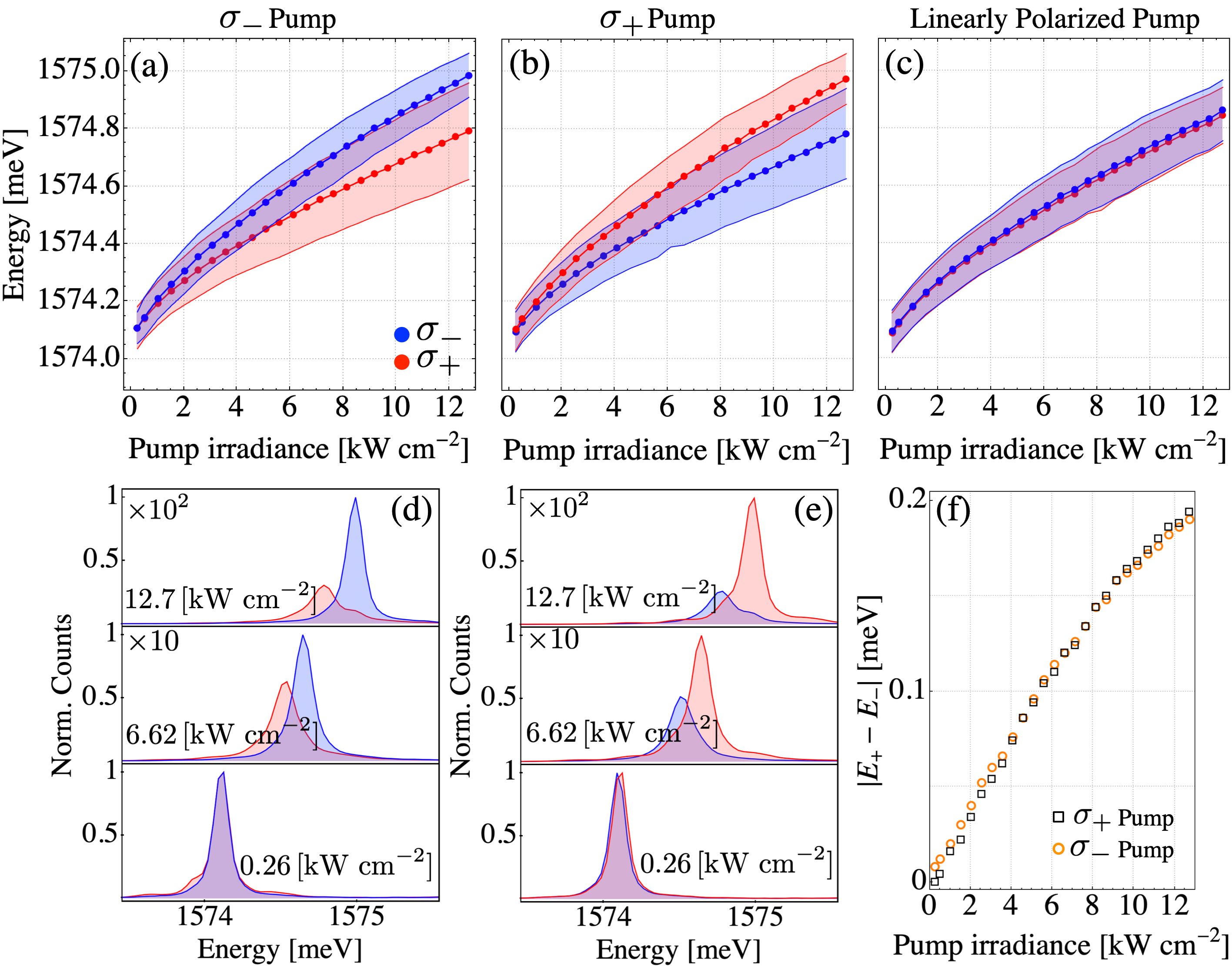}
  \caption{\label{fig2} Optical Zeeman effect in a micropillar. (a)-(c) Measured energy of \textit{s} orbitals as a function of the pump irradiance when detecting with circular polarization, $\sigma_-$ (blue dots) and $\sigma_+$ (red dots) with three different polarizations of the pump: (a) $\sigma_+$ circularly polarized, (b) $\sigma_-$ circularly polarized, (c) linearly polarized. Shadowed areas corresponds to the measured FWHM of the emission peak. (d),(e) Measured spectra at three pump irradiance values when pumping with $\sigma_-$ and $\sigma_+$ polarization, respectively. (f) Absolute value of the splitting between $\sigma_+$ and $\sigma_-$ peak energies as a function of the pump irradiance.
  	}
\end{center}
\end{figure}


To show this effect we start by focusing on the \textit{s} modes, for which there are two degenerate states at $E_s=1574.12$~meV with $\sigma_+$ and $\sigma_-$ polarizations. We consider a circularly polarized laser pump. Even though the pump is nonresonant, the polarization of the generated carriers is partially preserved and an excitonic spin imbalance is created with a majority spin determined by the polarization of the laser. This excitonic spin imbalance was reported in Ref.~\cite{CarlonZambon2019} under similar excitation conditions. At low pump power [0.26 KW cm$^{-2}$, Fig~\ref{fig2}(d)-(e)], interactions are negligible and the spin imbalance of the reservoir does not play any role: both $\sigma_+$ and $\sigma_-$ polaritons modes emit at the same energy. At high pump powers, excitons with spin $+$ interact mainly with $\sigma_+$ polaritons, while excitons with spin $-$ interact with $\sigma_-$ polaritons~\cite{Shelykh2004c}. As a consequence, when the pump irradiance is ramped up, polaritons co-polarized with the pump experience a higher energy blueshift than those cross-polarized. This is clearly evidenced in the spectral profiles displayed in Figs. \ref{fig2}(d) and \ref{fig2}(e). The peak energy of the $\sigma_+$ or $\sigma_-$ polarized polariton emission is shown in Fig.~\ref{fig2}(a) and (b) as a function of excitation density for $\sigma_-$ and $\sigma_+$ polarized pumps, respectively. The shadowed areas represent the FWHM (linewidth) of the emission peak for each polarization, which is about $70\,\mu$eV at very low pump irradiance. When the power is increased, not only a splitting appears between opposite polarizations, but also the polariton mode co-polarized with the pump exhibits a higher intensity and becomes narrower in linewidth due to the onset of stimulated relaxation from the reservoir to the highest occupied polariton mode at high powers. Note that under linearly polarized pump, the excitonic reservoir does not present any spin imbalance and both $\sigma_+$ and $\sigma_-$ polariton modes blueshift in the same way when the pump power is increased [see Fig.~\ref{fig2}(c)].

\begin{figure*}[t!]
\begin{center}
  \includegraphics[width=\textwidth]{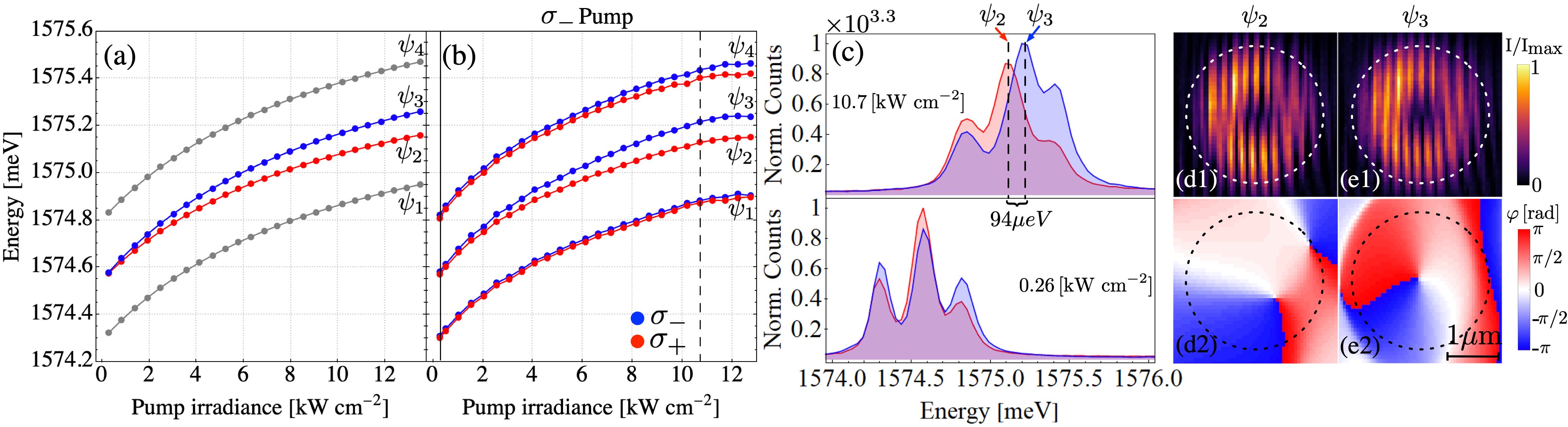}
  	\caption{\label{fig3} Chiral emission induced by optical Zeeman effect. (a) Computed energy of \textit{p} orbitals from Hamiltonian (\ref{Hami2}) as a function of excitation density (see text for parameters). (b) Measured energy of the \textit{p} modes as a function of the pump irradiance when detecting with the two circular polarizations. Blue (red) dots correspond to $\sigma_-$ ($\sigma_+$) polarization of emission, and the pump is $\sigma_-$ polarized. (c) Measured photoluminescence spectra at two values of the pump irradiance, which are denoted by continuous and dashed lines in (b). The optical Zeeman splitting between the $\psi_2$ and $\psi_3$ modes is highlighted in the top panel. (d1),(e1) Measured interference patterns of middle \textit{p} modes at the energies of the $\psi_2$ and $\psi_3$ modes, respectively, without any polarization selection. The orbital angular momentum of the modes is evidenced in (d2) and (e2), which show the retrieved phase gradients [see Fig.~\ref{fig1}(c) for comparison].}
   \end{center}
 \end{figure*}


The measured energy difference between the two opposite polarizations under circularly polarized pump is summarised in Fig.~\ref{fig2}(f) and shows a behaviour analogous to a Zeeman splitting in which the role of the external magnetic field is played by the power of the pump laser. The maximum observed splitting is $\sim 200\,\mu$eV at $P=12.7$~kW~cm$^{-2}$, which is of the same order than Zeeman splittings reported for exciton-polaritons with similar photon-exciton detuning, but under actual external magnetic fields of $4-9$~T in the Faraday configuration ~\cite{Larionov2010,Walker2011,Pietka2015,Sturm2015}. Increasing further the pump power does not enlarge the splitting because co-polarized polaritons enter into the lasing regime causing a saturation of their blueshift. Other power-dependent spin-relaxation effects may also be at the origin of the saturation~\cite{Dzhioev2002}. From the measured bleushifts and splittings, and assuming that only excitons and polaritons of the same spin interact, we estimate the ratio of $\sigma_-$ to $\sigma_+$ exciton populations to be $n_{+}/n_{-}=0.78$  at $P=12.7$~kW~cm$^{-2}$ under $\sigma_-$ pump.

The optically induced Zeeman splitting we have just described for the $s$ modes can be directly used to lift the degeneracy of middle modes in the $p$ multiplet, and to obtain the emission of modes with a net chirality. To experimentally demonstrate so, we now focus on power-dependent photoluminescence experiments in the \textit{p} modes. Taking advantage of the cavity wedge present in the wafer of the sample, we use a different micropillar of the same diameter with an emission energy for the \textit{p} modes very close to that of the \textit{s} modes discussed in Fig.~\ref{fig1} and, therefore, with the same photon-exciton detuning.

Figure~\ref{fig3}(b) displays the measured peak energy of each state extracted from a multi-gaussian fit as a function of the power irradiance of a $\sigma_-$ polarized pump when detecting either $\sigma_-$ or $\sigma_+$ polarizations. Every curve exhibits a monotonous blueshift of the energy when ramping up the power. More importantly, it is observed that modes $|\psi_3\rangle=|-1,\sigma_-\rangle$ and $|\psi_2\rangle=|+1,\sigma_+\rangle$ lift their degeneracy when the irradiance is increased. A maximum splitting $\Delta E_z=94\,\mu$eV is observed at  $P=10.7$~kW~cm$^{-2}$ [see Fig.~\ref{fig3}(c)]. This is in contrast with the lower and uppermost modes corresponding, respectively, to $\psi_1$ and $\psi_4$, which present a negligible polarization splitting (within our spectral resolution), see Fig.~\ref{fig3}(c).

The observed blueshifts and splittings in the \textit{p} manifold can be simulated using Eq.~(\ref{Hami2}) and considering a saturable exciton reservoir. We consider the total reservoir density as $n_{tot}=n_-+n_+=\beta P/(1+P/P_{sat})$, in which \textit{P} is the pump irradiance, $P_{sat}$ is the saturation value, and $\beta$ is a proportionality factor. Thus, the energy of the \textit{p} modes evolves as $E_p=E_{p0}+\alpha_1n_{tot}/2$, where now $E_{p0}=1574.58$ is the energy of the \textit{p} modes in absence of interactions and spin-orbit coupling, and $\alpha_1$ accounts for the interaction of excitons with same-spin polaritons (we neglect cross-spin interaction). Additionally, we consider the Zeeman term as $\Delta E_z=\alpha_1(n_--n_+)$, which takes into account the reservoir spin imbalance. Figure \ref{fig3}(a) shows the computed eigenenergies as a function of the pump irradiance \textit{P}, considering $n_+/n_-=0.86$, and $P_{sat}$ and $\alpha_1\beta$ as fitting parameters. They are in good agreement with the experimental values.
 
The direct consequence of the optically induced Zeeman splitting of states $\psi_2$ and $\psi_3$ is that orbital modes with opposite chirality are emitted at different energies. Therefore, frequency filtering of the emitted light is enough to select the mode with a given chirality. Figures~\ref{fig3}(d1)-(e1) show the interference pattern measured in the energy-resolved tomographic experiment described above at the energy of the $\psi_2$ and $\psi_3$ modes, respectively, at an excitation density of $P=10.7$~kW~cm$^{-2}$. In this experiment, the full emitted intensity at a given energy is measured without any polarization selection. From the interference patterns, the phase of the emitted field can be extracted by performing a filtered double Fourier transform, and it is shown in Fig.~\ref{fig3}(d2) and (e2) (see Ref.~\cite{Supplementary} for further details). A phase singularity is observed close to the center of the micropillars, with a clockwise phase winding of $2\pi$ around the center for the emission energy of $\psi_2$ and a counterclockwise winding at energy of $\psi_3$. The phase singularity visible at the upper right edge of the micropillar in Fig.~\ref{fig3}(d2) is an artifact related to a region of high gradient of intensity. Therefore, emission with energy-split vorticity is observed in our experiment thanks to the breaking of time-reversal symmetry optically induced by a polarized laser pump.

The method we have demonstrated to break time-reversal symmetry without the need of an external magnetic field has direct applications in the optical control of the topological phases of polariton Chern insulators in lattices~\cite{Bleu2016,Bleu2017b}. Indeed, the micropillar system we have discussed here is the building block to engineer one- and two-dimensional lattices. Moreover, the local control of the spin of the excitonic reservoir using external beams of different circular polarizations permits the breaking of time-reversal symmetry in more sophisticated manners. For example, one could envision lattices subject to staggered Zeeman fields, and interfaces between two regions of a lattice illuminated with beams of opposite circular polarizations showing a large number of topological edge modes. Our work has also interesting prospects in the manipulation of chiral modes at the single photon level~\cite{Kuriakose2021}. 

\noindent \textit{Acknowledgements}
This work was supported by European Research Council grants EmergenTopo (865151) and ARQADIA (949730), the H2020-FETFLAG project PhoQus (820392), the QUANTERA project Interpol (ANR-QUAN-0003-05), the Marie Sklodowska-Curie individual fellowship ToPol, the Paris Ile-de-France R\'egion in the framework of DIM SIRTEQ, the French National Research Agency project Quantum Fluids of Light (ANR-16-CE30-0021), the French RENATECH network, the French government through the Programme Investissement d’Avenir (I-SITE ULNE / ANR-16-IDEX-0004 ULNE) managed by the Agence Nationale de la Recherche, the Labex CEMPI (ANR-11-LABX-0007) and the CPER Photonics for Society P4S.

\bibliography{references}{}

\end{document}


\title{Supplemental Material: Chiral emission induced by optical Zeeman effect in polariton micropillars}%

\author{B.~Real}
\email{bastian.realelgueda@univ-lille.fr}
\affiliation{Univ. Lille, CNRS, UMR 8523 -- PhLAM -- Physique des Lasers Atomes et Mol\'ecules, F-59000 Lille, France}

\author{N.~Carlon Zambon}
\affiliation{Universit\'e Paris-Saclay, CNRS, Centre de Nanosciences et de Nanotechnologies, 91120, Palaiseau, France}

\author{P.~St-Jean}
\affiliation{Universit\'e Paris-Saclay, CNRS, Centre de Nanosciences et de Nanotechnologies, 91120, Palaiseau, France}

\author{I.~Sagnes}
\affiliation{Universit\'e Paris-Saclay, CNRS, Centre de Nanosciences et de Nanotechnologies, 91120, Palaiseau, France}

\author{A.~Lemaître}
\affiliation{Universit\'e Paris-Saclay, CNRS, Centre de Nanosciences et de Nanotechnologies, 91120, Palaiseau, France}

\author{L.~Le Gratiet}
\affiliation{Universit\'e Paris-Saclay, CNRS, Centre de Nanosciences et de Nanotechnologies, 91120, Palaiseau, France}

\author{A.~Harouri}
\affiliation{Universit\'e Paris-Saclay, CNRS, Centre de Nanosciences et de Nanotechnologies, 91120, Palaiseau, France}

\author{S.~Ravets}
\affiliation{Universit\'e Paris-Saclay, CNRS, Centre de Nanosciences et de Nanotechnologies, 91120, Palaiseau, France}

\author{J.~Bloch}
\affiliation{Universit\'e Paris-Saclay, CNRS, Centre de Nanosciences et de Nanotechnologies, 91120, Palaiseau, France}

\author{A.~Amo}
\affiliation{Univ. Lille, CNRS, UMR 8523 -- PhLAM -- Physique des Lasers Atomes et Mol\'ecules, F-59000 Lille, France}

\date{\today}%
\maketitle

\section{Setup and Experimental conditions}
\label{setup}
\begin{figure}[b]
\begin{center}
  \includegraphics[width=0.85\textwidth]{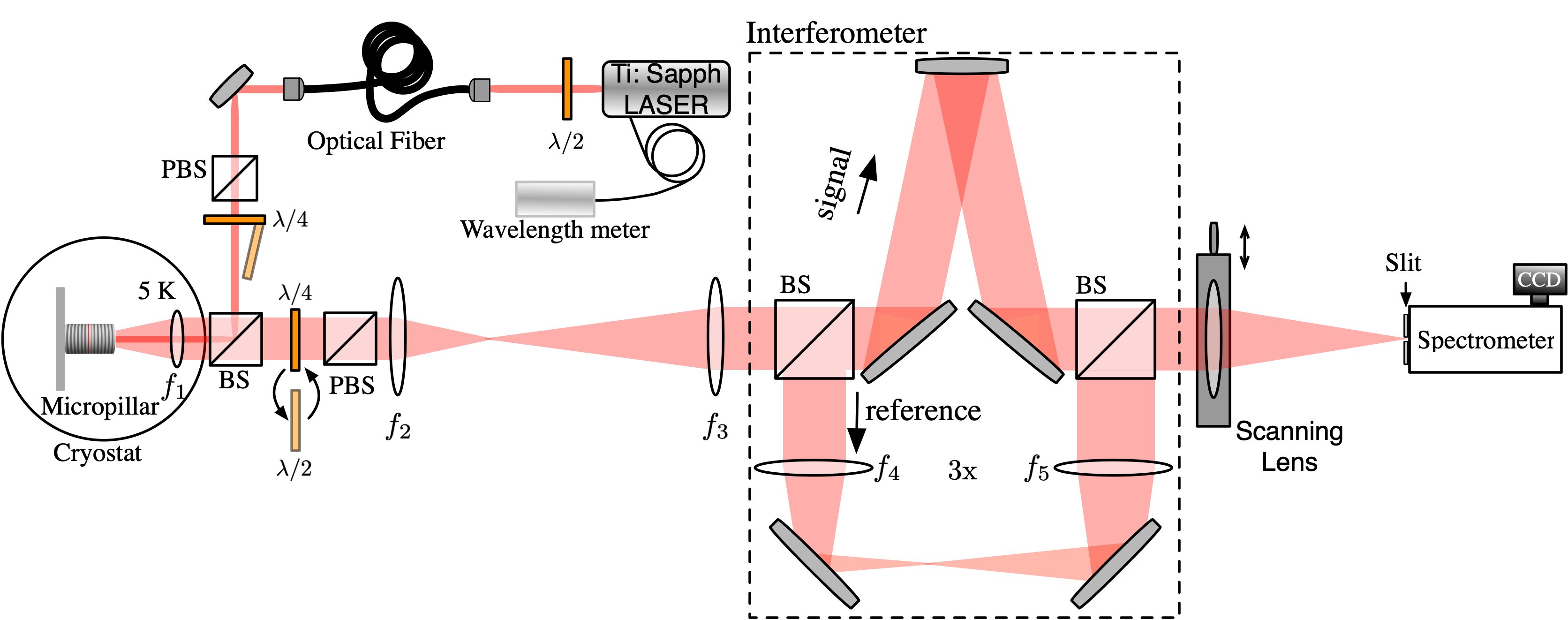}
  \caption{\label{figSuppl1} Scheme of the experimental setup to measure the self-interfered polariton emission. The region enclosed by a dashed rectangle highlights the interferometer along the detection path.} 
   \end{center}
 \end{figure}

The experiments shown in the main text are performed using the experimental setup schematized in Fig.~\ref{figSuppl1}. The sample is placed in a closed-cycle cryostat that allows to reach cryogenic temperature ($5$ K). We pump nonresonantly the micropillars by using a beam coming from a continuous-wave monomode Ti:Sapphire laser at $744.4$~nm. Before reaching the cryostat, the beam travels through a polarization-maintaining monomode optical fiber, which shapes it with a gaussian profile, and, then, it passes through a quarter-wave plate ($\lambda/4$) that sets its polarization (either linear or circular). The pump beams is finally focused on the micropillars by an aspherical lens with $8$-mm focal length (NA$=0.5$) that is placed inside the cryostat [lens $f_1$ in Fig.~\ref{figSuppl1}]. The full width at half maximum (FWHM) of the beam is $2.5\,\mu$m on the sample.

When the pump beams reaches the micropillar, photons are absorbed and a gas of excitons is created. Then, this excitonic reservoir relaxes and populates the polaritonic modes of the micropillar. Since polaritons have a finite lifetime they leak out of the micropillar in the form of photons that encode the energy and momentum of the polaritons inside the structure. The emission is collected by the same $8$-mm focal lens ($f_1$) and then magnified and imaged at the entrance slit of a spectrometer. The entrance slit together with a scanning lens placed on a motorized translation stage allow selecting real-space vertical slices of the micropillar emission. Each slice of the image is dispersed by the spectrometer and imaged on a Coupled-Charge Device (CCD) camera. By recording these spectra, the real-space pattern of the different modes of the micropillar can be reconstructed with an energy precision of $33.1\,\mu$eV. Furthermore, polarization-resolved spectroscopy can be done by placing a quarter-wave plate ($\lambda/4$) or a half-wave plate ($\lambda/2$) (circularly or linearly polarized detection, respectively) and a polarized beam splitter (PBS) along the detection path.

To measure the phase of \textit{p} modes of the micropillar, an interferometer is set up along the detection path (see enclosed region by a dashed rectangle in Fig.~\ref{figSuppl1}). Here, the detection path is split into two paths of the same length (zero delay time between the two arms). Along one of the arms, the image of the micropillar is magnified by a factor of 3 (reference beam) with respect to the image obtained along the other arm (signal beam). Consequently, the interference of the emission travelling throught the two arms can be recorded on the CCD camera, allowing the reconstruction of an interference pattern at the energy of each mode [see Fig.~\ref{figSuppl4}(a)].    
\section{Linearly polarized measurements and Stokes Parameters}
\begin{figure}[h!]
\begin{center}
  \includegraphics[width=0.77\textwidth]{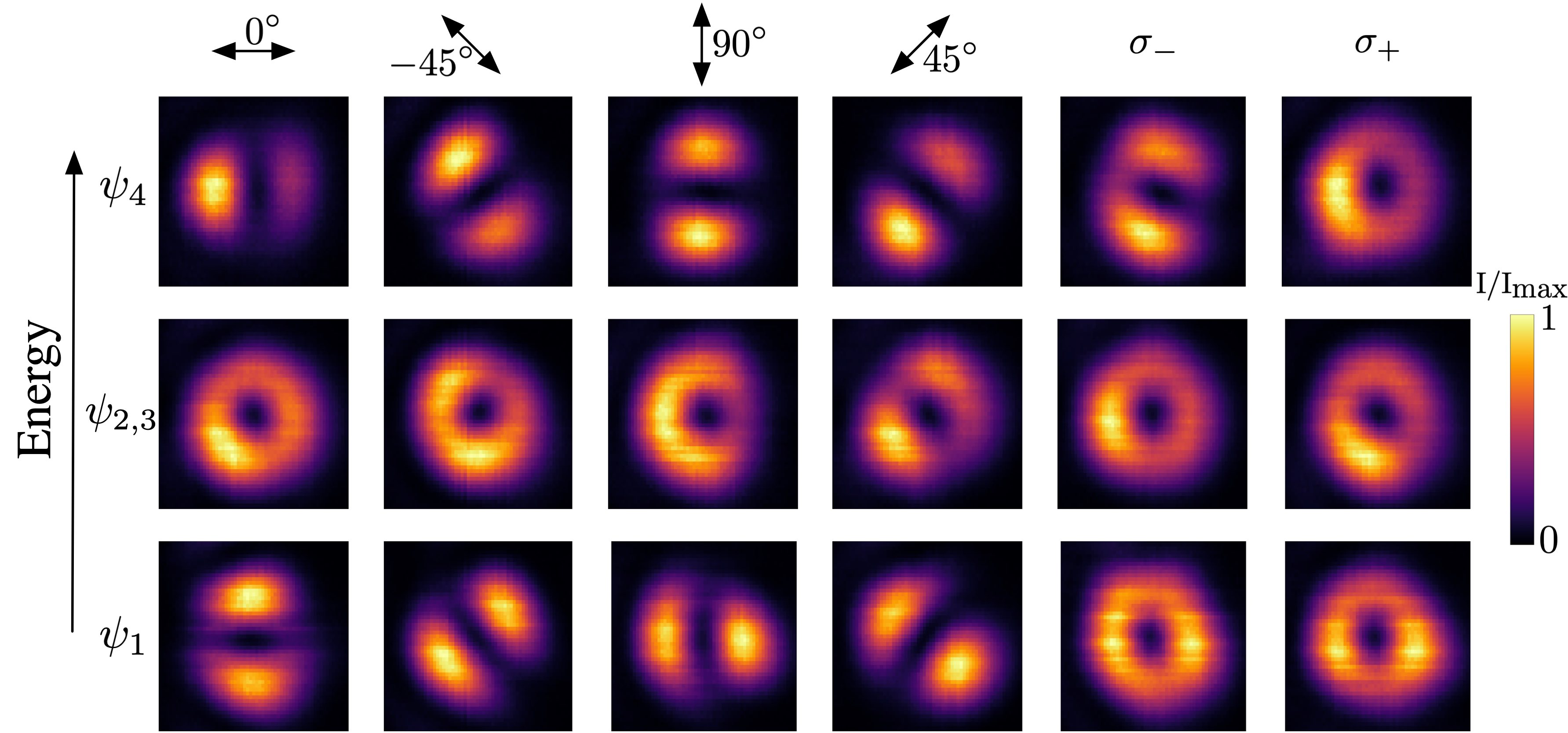}
  \caption{\label{figSuppl2}Tomographic images of the \textit{p} modes with different linear and circular polarization pointed out on top. Each images is normalized to its maximum. } 
   \end{center}
 \end{figure}

To obtain the polarization textures of the \textit{p} modes presented in Fig.~1(c) of the main text, polarization-resolved tomography is carried out at low pump irradiance ($P=0.26$~kW~cm$^{-2}$). Figure~\ref{figSuppl2} shows the reconstructed images at the energy of the three emission lines of the \textit{p} modes when selecting linear polarization along horizontal ($0^{\circ}$), vertical ($90^{\circ}$), diagonal ($45^{\circ}$) and antidiagonal ($-45^{\circ}$) directions, and circular polarization ($\sigma_+$ and $\sigma_-$). 

\begin{figure}[h!]
\begin{center}
  \includegraphics[width=0.52\textwidth]{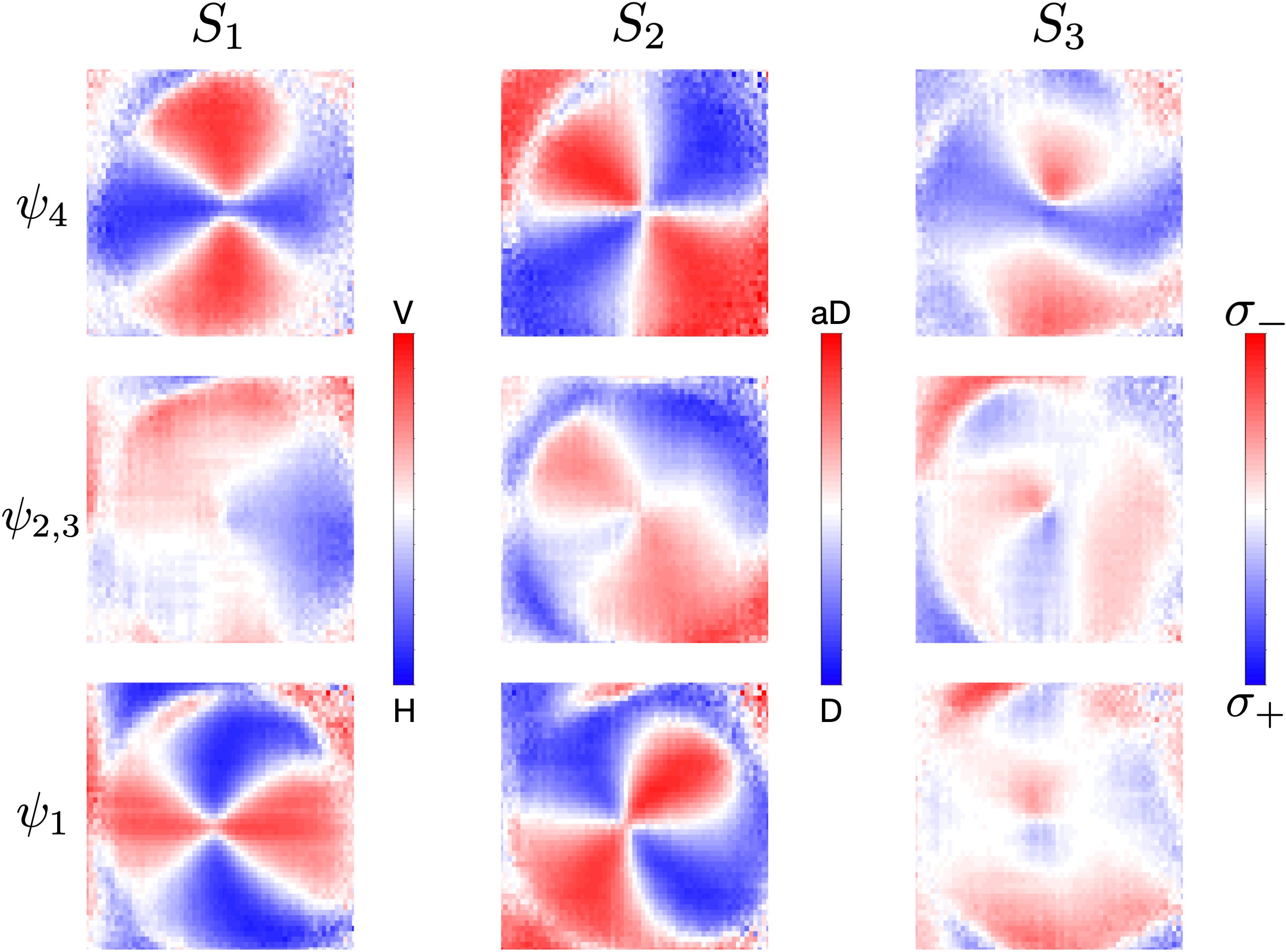}
  \caption{\label{figSuppl3} Stokes parameters $S_1$, $S_2$ and $S_3$ for $\psi_1$, $\psi_2$, $\psi_3$ and $\psi_4$ \textit{p} modes introduced in the main text. } 
   \end{center}
 \end{figure}

 From these images, the Stokes parameters for each pixel pixel in real space at a given energy computed. These parameters are defined as: 
\begin{equation}
S_1=\frac{I_H-I_V}{I_H+I_V}\,,\;S_2=\frac{I_{45^{\circ}}-I_{-45^{\circ}}}{I_{45^{\circ}}+I_{-45^{\circ}}}\,,\;S_3=\frac{I_{\sigma_+}-I_{\sigma_-}}{I_{\sigma_+}+I_{\sigma_-}}\,.
\end{equation}
Figure \ref{figSuppl3} shows the Stokes parameters for the three emission energies of the \textit{p} modes. For $\psi_1$ and $\psi_4$, the $S_3$ Stokes parameter has very weak values (mostly light-blue and light-red colors), while $S_1$ and $S_2$ reach values close to $1$ and $-1$ (dark red/blue, respectively). This means that these modes are well caracterized by the linear polarization angle $\phi=(1/2)\arctan(S_2/S_1)$~\cite{Dufferwiel2015}. Figure~1(c) of the main text shows the otientation of the linear polarization at each point in space extracted from this angle.

\newpage 
\section{Phase retrieving}
\begin{figure}[h!]
\begin{center}
  \includegraphics[width=0.7\textwidth]{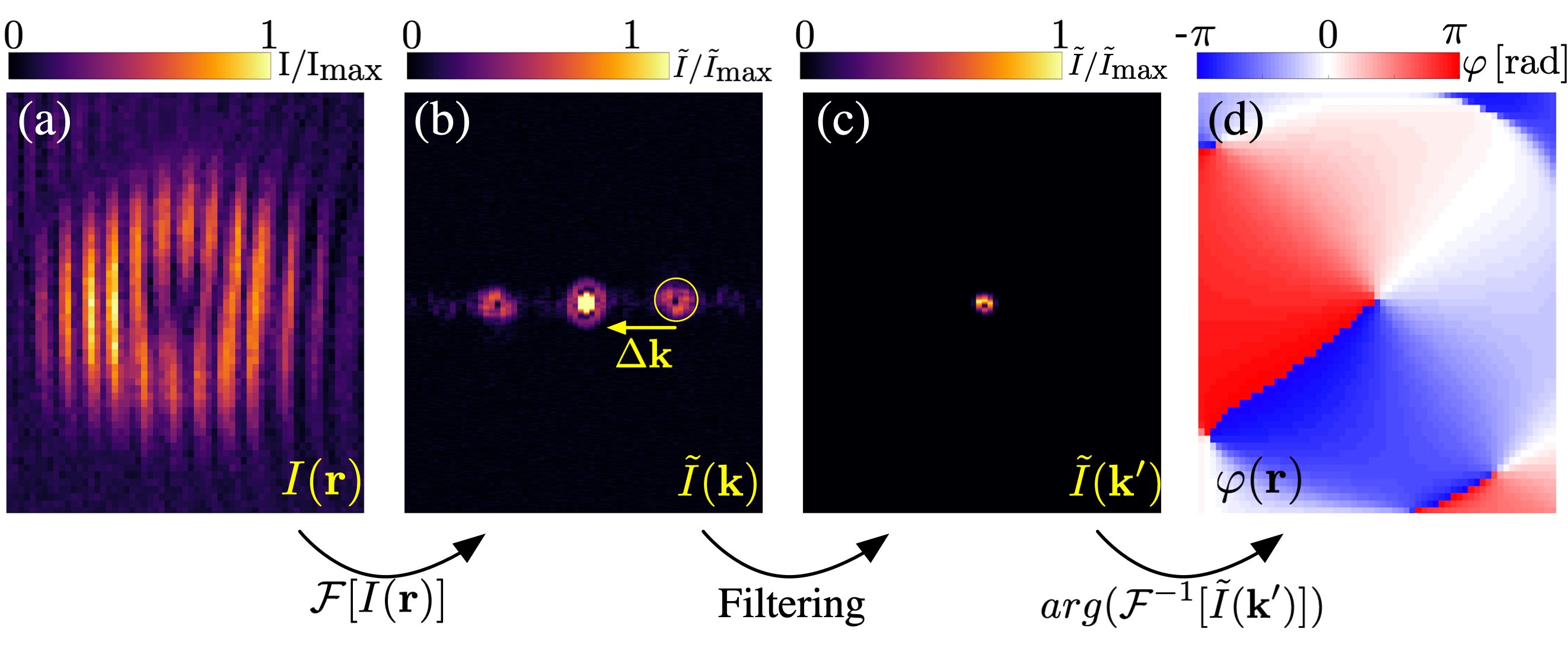}
  \caption{\label{figSuppl4}  Retrieving of the wavefront phase. (a) Interference pattern for the middle \textit{p} mode when selecting $\sigma_-$ polarization at low pump irradiance [see Fig.~1(c) in the main text]. This is produced by interfering the signal and reference arm (see Fig.~\ref{figSuppl1}) with a given wavevector difference $\Delta {\bf k}$. (b) Fast Fourier transform (FFT) of the interferogram. (c) Filtered and translated image of (b). (d) Real-space phase pattern $\varphi({\bf r})$ of the emission.} 
   \end{center}
 \end{figure}

Figures 1 and 3 of the main text show the retrieved phase of the middle \textit{p} modes when selecting the circular polarization of emission [Fig.~1(c)] and the energy [Fig.~3(d1),(e1)]. In order to obtain these images, a Fourier filtering procedure of the interference pattern was implemented, which is described below. Figure~\ref{figSuppl4}(a) shows the measured image at the entrance slit of the spectrometer for $|\psi_3\rangle$ mode ($\sigma_-$ polarized detection) when the signal is split into two optical beams as described in the setup above. These two beams can be written as:

\begin{eqnarray*}
{\bf A}_s({\bf r},t)&=& {\bf A}_s({\bf r})e^{-i(\omega t -{\bf k}_s\cdot {\bf r} +\varphi({\bf r}))}\,,\\
{\bf A}_r({\bf r},t)&=& {\bf A}_r({\bf r})e^{-i(\omega t -{\bf k}_r\cdot {\bf r})}\,,
\end{eqnarray*}      
\noindent where ${\bf A}_s({\bf r})$ (${\bf A}_s({\bf r})$) is the amplitude of the signal (reference) beam, ${\bf r}=(x,y)$ is the real-space vector across the transverse plane of the micropillar (perpendicular to the growth direction $z$), $\omega$ is the frequency of the emitted mode (selected by the spectrometer), ${\bf k}_s$ and ${\bf k}_r$ are the transverse wavevectors of the signal and reference beam, respectively, and $\varphi(\bf r)$ is the wavefront phase of the \textit{p} mode that varies throughout the ${\bf r}$ plane of the micropillar. Since the reference beam is three times bigger than the signal one, we assume the reference beam having a constant wavefront phase in the region of interference with the signal beam. Therefore, the interference can be written as:

\begin{eqnarray}
I({\bf r})&=&|{\bf A}_s({\bf r},t)+{\bf A}_r({\bf r},t)|^2\,,\nonumber\\
I({\bf r})&=&|{\bf A}_s({\bf r})|^2+|{\bf A}_r({\bf r})|^2 + \left({\bf A}_s({\bf r}){\bf A}_r^*({\bf r})e^{-i(\Delta {\bf k\cdot {\bf r}} + \varphi({\bf r}))}+cc.\right)\,,
\label{EqInterf}
\end{eqnarray}

\noindent where $\Delta {\bf k}={\bf k}_s-{\bf k}_r$ is set by the angle between the signal and reference beams at the entrance slit of the spectrometer. The angle between the two beams lies within the horizontal plane, realusting in an interference fringes in the vertical direction. The Fourier transform of this interference pattern, $\tilde{I}({\bf k})=\mathcal{F}[I({\bf r})]$, gives three peaks that are shown in Fig.~\ref{figSuppl4}(b). One peak is at the center of the reciprocal space and it comes from the first and second terms of eq.~(\ref{EqInterf}). The other two peaks are translated by $\pm \Delta {\bf k}$ with respect to the origin and they come from the third and fourth terms of eq.~(\ref{EqInterf}). In consequence, these non-centered peaks carry the information of the wavefront phase $\varphi({\bf r})$. In order to retrieve it, firstly, a translation equal to $-\Delta {\bf k}$ of the reciprocal space is done, thus, the right peak stays on the center of the reciprocal space. Then, a filtering process erases all other frequencies as Fig.~\ref{figSuppl4}(c) shows. Specifically, this filtering process selects the frequencies only within a small region in the center by using a circular (sharp) or gaussian-like shape, giving both similar results. The remaining peak corresponds to $\tilde{I}({\bf k'})=\mathcal{F}[{\bf A}_s({\bf r}){\bf A}^*_r({\bf r})e^{-i\varphi({\bf r})}]$. Finally, an inverse Fourier transform of the filtered momentum space image is implemented. Its argument directly gives the phase $\varphi({\bf r})$ shown in Fig.~\ref{figSuppl4}(d).      

\bibliography{referencesSuppl}{}